%% file: main.tex
\pdfoutput=1
\documentclass[10pt]{article}
\input{preamble.tex}

\title{Does Your Agent Need to Know Its Budget? A Placebo-Controlled Experiment on LLM Coding Agents}

\author{Iftesam Nabi\textsuperscript{1} \quad Jarif Ul Alam\textsuperscript{2} \quad Sunzid Hassan\textsuperscript{1} \quad Sabbir Alom Shuvo\textsuperscript{1} \\[0.2em]
 Md Faiaz Al Islam\textsuperscript{1} \quad Yead Rahman\textsuperscript{1} \quad Mehrab Hosain\textsuperscript{1} \\[0.8em]
 \textsuperscript{1}College of Engineering and Science, Louisiana Tech University, Ruston, LA 71272, USA \\
 \textsuperscript{2}College of Business, Louisiana Tech University, Ruston, LA 71272, USA \\[0.4em]
 {\small ina003@latech.edu, jua003@latech.edu, sha040@latech.edu,} \\
 {\small sas115@latech.edu, mfa019@latech.edu, yead1409031@gmail.com, mho020@latech.edu}}

\date{}

\begin{document}
\maketitle
\begin{abstract}
\itshape
Organizations deploying autonomous coding agents impose token budgets, yet the causal effects of a budget, and of telling the agent about it, are unmeasured. We conduct a placebo-controlled factorial experiment under a frozen, timestamped protocol, crossing budget enforcement with budget disclosure on two executable program-repair benchmarks (5{,}835 runs, two open models, three cap levels). Enforcing a calibrated cap reduced task resolution by 20.0 percentage points while reducing tokens per resolved task roughly twelvefold. Disclosing the enforced budget never helped: the placebo-controlled risk difference was $-2.8$\,pp (95\% CI $[-6.2, +0.7]$) at the medium cap and $-4.2$\,pp ($[-8.1, -0.2]$, nominally significant, unadjusted) at the tight cap, with every prompt wording and per-turn feedback variant far below the 10-point effect size of interest. Observed run-to-run outcome variance was roughly 1.5$\times$ the uncapped level under caps, though part of this reflects resolution rates moving nearer one half. These results make budget governance for coding agents an enforcement-design problem: no communication variant produced an effect beyond that of an unrelated sentence, which itself raised resolution by 7.2\,pp (95\% CI $[+4.0, +10.5]$) over silent enforcement, a sentence effect that only a placebo control removes.
\end{abstract}

\vspace{1em}

\section{Introduction}

Large language model (LLM) agents now iterate through reasoning, tool use, and code edits~\parencite{Yang2024SWEagent} to autonomously resolve issues drawn from real software repositories~\parencite{Jimenez2024SWEbench}. Their expenditure is large and erratic: an agentic coding task consumes roughly a thousandfold more tokens than a conversational coding exchange, and identical tasks vary up to thirtyfold in tokens across runs~\parencite{Bai2026Spend}. Organizations therefore impose budgets, typically as termination caps that end a run once spending exceeds a limit~\parencite{Yang2024SWEagent, Liu2024AgentBench}. A cap forces two policy choices: where to set the limit, and whether to tell the agent it exists. Both choices are currently made without causal evidence: evaluation practice reports cost beside accuracy, but what an enforced budget, disclosed or silent, does to task resolution remains unmeasured~\parencite{Yang2024SWEbenchM, Fan2025SWEEffi}.

Prior work has made agent expenditure visible and bounded through three families of controls. Cost-aware evaluation reports dollar cost beside resolve rate, scores systems by resolution achieved under token and dollar budgets, and measures the run-to-run consistency of resource use~\parencite{Yang2024SWEbenchM, Fan2025SWEEffi, Rabanser2026Reliability}. Structural designs bound spending at the workflow level: replacing open-ended tool use with a fixed pipeline cut per-issue cost from roughly \$2.50 to \$0.70 while resolving more issues~\parencite{Xia2024Agentless}. Allocation policies act at dispatch: budget-constrained routing retains 93\% of frontier quality at a quarter of the cost, and adaptive test-time compute matches fixed sampling with a quarter of the samples~\parencite{Panda2025Routing, Snell2024TestTime}. A separate behavioral literature suggests the communication of a limit is itself a lever: specific, difficult targets reliably outperform open-ended instruction when performers receive feedback on progress~\parencite{Locke2002Goal, Locke2006Goal}. These strands measure, design, and allocate, yet none isolates what an enforced budget, or knowledge of it, causes an agent to do on otherwise identical tasks.

A central limitation is that in deployed systems the budget appears only as a terminal event: a cap that ends the run~\parencite{Yang2024SWEagent, Liu2024AgentBench}. Breached caps are routine: cost-limit or error terminations and exhausted round allowances are among the most common outcomes in agent evaluations~\parencite{Yang2024SWEbenchM, Liu2024AgentBench}. Cost tables built from such runs cannot say what the cap caused: expenditure and outcome are produced jointly by the trajectory, so a cheap failure is indistinguishable from a failure the cap itself created. The confound extends to communication: systems that enforce limits may also disclose them, so enforcement and disclosure arrive entangled at any given limit. Underneath sits the assumption that a budget is an accounting parameter: on this view, its effects can be read off observed spending and need no intervention to estimate. The unresolved problem, therefore, is causal identification: what an enforced budget, and separately its disclosure, does to resolution, to expenditure, and to reliability across runs.

Separating these effects requires holding the limit fixed while varying only whether the agent is told of it. The behavioral evidence gives the disclosure arm a direction: performers given explicit targets regulate effort toward them when they can track progress, so an informed agent should waste less of a binding budget~\parencite{Locke2002Goal, Locke2006Goal}. The transfer is analogical: goal-setting evidence concerns performance targets, while a budget is a resource constraint, so treating the cap as a specific, difficult target is itself a premise the experiment puts under test. Therefore, we hypothesize that announcing an enforced token budget yields equal or higher task resolution than enforcing the same budget silently, while reducing wasted expenditure. Accordingly, this paper asks whether budget communication, at an identical enforced cap, changes resolution probability, token expenditure, and run-to-run stability in coding agents. If it does, the cheapest control an organization owns, a sentence in a prompt, becomes a governance instrument; if it does not, only enforcement design matters.

To answer this question, we conduct a placebo-controlled factorial experiment, under a frozen and timestamped protocol, that crosses budget enforcement with budget disclosure at caps calibrated from held-out development tasks, with repeated runs per task and condition. Effectiveness is evaluated by measuring harness-verified resolution rate, billed tokens per resolved task, and the variance of outcomes across repeated runs, on two executable program-repair benchmarks. The placebo condition, a token-matched neutral sentence at the same enforced cap, separates the effect of budget semantics from the effect of adding any instruction at all. To our knowledge, this is the first randomized, placebo-controlled estimate of budget-enforcement and budget-disclosure effects for coding agents. The placebo cell also carries a finding of its own: because an added sentence of any content shifted outcomes, prompt-level interventions evaluated against an empty prompt inherit a sentence effect that only a placebo control removes. By estimating enforcement and disclosure effects separately, this study aims to give organizations evidence for setting and communicating agent budgets, and the field a replicable template for governance experiments.

\section{Related Work}

Autonomous coding agents are evaluated along three coupled dimensions: whether they resolve the issue, what resources they consume doing so, and how consistently they behave across runs~\parencite{Jimenez2024SWEbench, Fan2025SWEEffi, Rabanser2026Reliability}. These dimensions conflict: additional expenditure buys resolution only up to a point, with accuracy peaking at intermediate cost and saturating beyond it~\parencite{Bai2026Spend}. Leaderboard practice nevertheless ranks systems by resolve rate alone, treating resources as effectively unlimited~\parencite{Fan2025SWEEffi}. The resulting risk is that deployment decisions about spending are made on evidence that never measured spending's effects.

Existing responses fall into three families: interactive agent scaffolds with built-in cost limits, efficiency-oriented workflow designs, and budget-aware allocation policies. Among scaffolds, SWE-agent pairs a language model with an agent-computer interface and enforces a per-instance dollar budget, submitting existing edits automatically when a run exceeds it~\parencite{Yang2024SWEagent}. Efficiency-oriented designs restrict the agent instead: Agentless replaces open-ended tool use with a fixed localization, repair, and validation pipeline, resolving more issues than agentic baselines at \$0.70 per issue~\parencite{Xia2024Agentless}. ReAct caps trajectories at five to seven steps after finding that more steps do not improve performance, and Toolformer permits at most one tool call per input to prevent unproductive calling loops~\parencite{Yao2023ReAct, Schick2023Toolformer}. Allocation policies act across queries: cost cascades route queries through successively cheaper models~\parencite{Chen2023FrugalGPT}, budget-constrained routing with PILOT retains 93\% of GPT-4 quality at a quarter of its cost, and compute-optimal test-time scaling matches best-of-N sampling with a quarter of the samples~\parencite{Panda2025Routing, Snell2024TestTime}.

Despite this progress, every family treats the budget as a configuration constant; none treats it as a variable under study. Benchmarks record cost after the fact, scaffolds fix caps as evaluation conveniences, and routing allocates before the trajectory begins, so the cost-accuracy tables the field produces are observational~\parencite{Fan2025SWEEffi, Yang2024SWEbenchM}. Even resolution-under-budget scoring evaluates spending without intervening on policy: no cost-aware evaluation randomizes the budget or its disclosure. Expenditure and outcome are generated jointly by the same trajectory, and observational tables cannot separate what the budget caused from what the task and model would have produced anyway. As a result, existing approaches optimize aggregate cost-effectiveness without encoding budget policy (its level or its communication) as a manipulated treatment.

This limitation becomes consequential because budgets bind routinely: nearly half the runs of some configurations end at the cost limit or an error, and exhausted round allowances are on average the most common failure elsewhere~\parencite{Yang2024SWEbenchM, Liu2024AgentBench}. The spending being governed is itself unpredictable: token consumption disperses widely across repeated runs of the same task, and models systematically underestimate their own future usage~\parencite{Bai2026Spend}. That underestimation mirrors the planning fallacy documented in human forecasters, who understate time and cost even when underestimation is penalized~\parencite{Kahneman1977Intuitive}. Aggregate cost-accuracy figures therefore mask the local damage a mis-set or mis-communicated cap can do: runs stopped before solutions they might have reached, and failure misattributed to the model.

The field traditionally fixes both choices, the cap and its communication, as harness constants and ad hoc prompt conventions, chosen before any run and never varied within a study~\parencite{Yang2024SWEagent, Xia2024Agentless}. An efficient-reasoning literature does manipulate budget language directly: stating a numeric token budget in the prompt compresses chain-of-thought at near-constant accuracy, and test-time scaling work controls reasoning length by forcing generation to a stated budget~\parencite{Han2024TALE, Nayab2024CCoT, Muennighoff2025s1}. These studies measure length compliance in single-turn reasoning, without runtime enforcement, an executable task outcome, or a placebo control; whether a stated budget changes what a multi-turn agent resolves under an enforced cap is untested there. Behavioral research suggests the choice is not neutral either: specific, difficult targets outperform open-ended instruction when performers receive progress feedback, and resource-rational analysis models effective bounded agents as weighing each computation's expected benefit against its cost~\parencite{Locke2002Goal, Locke2006Goal, Griffiths2015Rational, Lieder2020Resource}. Whether either mechanism transfers to language-model agents has not been tested by intervention. Our work removes this fixed-compromise design: budget enforcement and budget disclosure become separately randomized treatments in a preregistered, placebo-controlled factorial experiment that holds the cap fixed while varying only the agent's knowledge of it.

Taken together, the literature shows clear progress in measuring, engineering, and allocating agent expenditure, but leaves open what an enforced budget, or the agent's knowledge of one, causes. This motivates methods that move beyond post hoc cost accounting and instead intervene on budget policy under controlled conditions. Consequently, effectiveness must be measured through a tripartite evaluation: harness-verified resolution probability, billed tokens per resolved task, and the stability of outcomes across repeated runs.

\section{Research Questions and Hypotheses}

The study asks four questions, registered in a protocol frozen and timestamped on 2026-08-05, before confirmatory data collection; the protocol was not deposited in an external registry, and the frozen document with its dated amendment log ships in the study archive. The term preregistered is used below in this internal-registration sense. RQ1 concerns the frontier: how an enforceable token budget affects resolution probability and token expenditure. Its hypothesis, H1, expects tighter budgets to lower expected resolution, with magnitude varying by task and model. RQ2 concerns disclosure and carries the headline hypothesis: at an identical enforced cap, does communicating the budget change resolution or resource waste relative to silent enforcement? H2 expects announced caps to yield equal or higher resolution than silent caps while reducing wasted expenditure. RQ3 asks whether budget sensitivity differs across models and task pools, and RQ4 asks whether caps and their communication change the variance of outcomes across repeated runs. The primary preregistered estimand is the difference in resolution probability between the announced and placebo cells at the medium cap; every other comparison is secondary.

\section{Methodology}

The study uses a randomized factorial intervention: budget enforcement and budget disclosure are manipulated as crossed treatments over a fixed task set. Randomization is required because expenditure and outcome are produced jointly by each trajectory, so observational cost accounting cannot identify the effect of either policy. A placebo condition is included because an added prompt sentence may alter behavior independently of its budget content, which would confound disclosure with prompt modification. Repeated runs per task and condition are included because agent outcomes vary across runs; single-run comparisons are therefore uninformative about either effects or stability.

Two task pools were used. Pool A comprises 29 single-defect Python programs with input-output test cases from QuixBugs~\parencite{Lin2017QuixBugs}; two of 31 candidates were excluded by a frozen rule because their reference implementations exceed the 20-second evaluation timeout. Pool B comprises the 164-task Python split of HumanEvalFix, in which human-written buggy solutions are paired with hidden unit tests~\parencite{Muennighoff2023OctoPack}. Each pool was split into development and confirmatory sets by a deterministic index rule frozen before any model run: alphabetical index modulo three equal to zero selects Pool A development tasks, task index modulo six equal to zero selects Pool B development tasks, and the two timeout exclusions were removed after the split without reshuffling. The resulting splits are 11 development and 18 confirmatory tasks for Pool A, and 28 and 136 for Pool B. Every task was validated before use by requiring its reference program to pass all tests and its buggy program to fail at least one.

The harness consists of three components: a repair loop, an execution evaluator, and a budget controller. The repair loop sends the buggy program and test feedback to the model, and receives one complete candidate program per turn. Feedback reports only the number of hidden tests passed, never failing inputs or expected outputs. The execution evaluator runs each candidate in a separate process with a 20-second limit and returns the test summary that becomes the next turn's feedback. The budget controller maintains a per-run ledger in billed-equivalent tokens, defined per call as context length plus generated tokens, with context reconstructed exactly from the runtime's evaluated-token counts under prefix caching. Models were served by Ollama 0.32.1 on CPU; the primary model is qwen3-coder-30B (Q4\_K\_M quantization) and the secondary model is qwen3-4B with extended thinking disabled.

Development tasks were first run uncapped, subject only to an invisible 40{,}000-token safety ceiling, with five seeds in Pool A and three in Pool B to calibrate cap levels. The preregistered percentile cap rule degenerated at development time, because seven of eleven Pool A development tasks could not fund a first call under its medium cap; a logged amendment replaced it before any confirmatory result. The replacement freezes caps at one, two, and four multiples of a unit attempt cost (tight, medium, and generous). The unit cost equals the 75th percentile of development first-call billed tokens, rounded to the nearest 50 and raised at least to the maximum first-call requirement, context plus the 512-token reserve. This gave 924, 1{,}848, and 3{,}696 tokens for Pool A and 766, 1{,}532, and 3{,}064 for Pool B. A second logged amendment raised the Pool B tight cap to 1{,}000, because 766 left 24 of 164 pool tasks unable to initiate a first call. Each run used temperature 0.7, a fixed per-run seed, 512 maximum output tokens, a 16{,}384-token context window, at most 10 turns, a 40-minute wall-clock limit, and termination after three replies without a code block. A call was blocked, and the run terminated as budget-exhausted, whenever accumulated billed tokens plus the projected next context (the known cached context plus incoming message characters divided by 3.5) and the 512-token output reserve exceeded the active cap. A response that landed past the cap likewise ended the run, without that turn's candidate being evaluated. The full study comprises 5{,}835 runs across calibration, confirmatory, and supplementary arms; post-calibration arms total 5{,}666. No post-calibration run ended in a harness error, so none was excluded. Confirmatory runs used identical per-condition seed lists drawn from seeds 1--4, four seeds in the medium-cap factorials and two or three in supplementary arms. They executed within-task round-robin on a 32-core, 256~GB CPU virtual machine over three days; sampling was seeded but not bitwise deterministic.

Five conditions were compared at the medium cap, selected to cross the statement of a budget with its enforcement. Baseline (no budget sentence, invisible ceiling) establishes uncapped performance; silent-bare (enforced cap, no sentence) isolates enforcement without communication. Info-only states the medium cap without enforcing it, isolating the statement itself, and announced states the cap and enforces it. Silent-placebo adds a content-neutral sentence of matched length at the same enforced cap; it is the primary comparator for announced because it separates budget semantics from the effect of adding any sentence. The frozen wordings, appended to the system prompt, are: budget sentence, ``This task has a total budget of \{cap\} tokens; all input and output tokens count toward it.''; placebo, ``This task is one item in a batch of automated program-repair evaluations run on this system.'' The countdown variant injects per turn ``Budget remaining: \{remaining\} of \{cap\} tokens.'' and its control injects ``Automated evaluation system message: attempt log entry recorded.''; the plan and consequence wording variants append ``Plan your work and stay within this budget.'' and ``If the budget is exceeded, the attempt ends immediately and is scored as-is.'', and the effective-budget variant states the cap minus the 512-token reserve. Supplementary conditions repeated the contrast at the tight cap and added countdown, neutral-injection, wording, second-model, and identifier-renamed variants. All conditions shared tasks, seed lists, sampling parameters, and evaluator. Three logged amendments changed shared parameters: lean test feedback together with a per-call output reserve reduced from 1{,}024 to 512 tokens, both adopted after a rich-feedback pilot resolved 26 of 27 development runs mostly in a single call; the attempt-based cap rule; and the Pool B tight-cap raise. A fourth, larger amendment preceded these: the registration originally reserved confirmatory hypothesis testing for a cloud-hosted frontier-model arm on repository-scale issues, and when no cloud budget was available, a timestamped amendment (2026-08-05, before any Pool B model run) added Pool B, designated its medium-cap factorial as the confirmatory test, and re-labeled the cloud arm as a replication package. Each parameter amendment was adopted before any affected confirmatory outcome was examined; the eight runs already launched under superseded settings were discarded unexamined. The term confirmatory is reserved for the Pool B medium-cap factorial: the tight-cap contrast was a powered follow-up, and the countdown, wording, second-model, and perturbation arms were registered as exploratory.

The primary outcome is the resolved rate: the proportion of runs in which an evaluated candidate passes every task test, with higher values indicating better resolution. The frozen analysis plan specified a hierarchical logistic regression with task random intercepts as primary, with task-stratified permutation inference and task-cluster bootstrap intervals as its designated fallback chain; the fallback chain produced all reported statistics, and the hierarchical model was not estimated. Because every cell runs the identical task-by-seed grid, the unadjusted risk difference coincides with the frozen g-computation estimand over the observed tasks. Condition contrasts are reported as risk differences with 95\% percentile intervals from 5{,}000 bootstrap resamples of tasks as clusters~\parencite{Efron1979}, keeping all of a resampled task's runs. Significance is assessed by two-sided permutation tests using 5{,}000 within-task label permutations at $\alpha = .05$; 10 percentage points was preregistered as the smallest effect size of interest. The registration's power simulation indicated at least 80\% power to detect a 10-point disclosure effect on the 136-task, four-seed factorial, and the realized interval half-width of 3.5 points confirms the design resolved effects well below that threshold. Only the announced versus silent-placebo contrast at the medium cap is confirmatory; all other contrasts are secondary, reported without multiplicity adjustment, and of their $p$-values only the enforcement and placebo-sentence contrasts (both $p=.0002$) would survive a Bonferroni correction across the four secondary tests. Expenditure is summarized as tokens per resolved task, computed as the sum of billed tokens across a condition's runs divided by its resolved count; lower values indicate greater efficiency. Stability is measured as the mean per-task population variance (denominator $n$) of the binary outcome across seeds within a condition; lower values indicate more repeatable behavior. Because the variance of a binary outcome is bounded by $p(1-p)$ and peaks at one half, this statistic partly reflects mean shifts and is reported descriptively. Memorization is screened as the normalized string similarity between a signature-only completion and the reference solution, with tasks above 0.9 flagged; termination reasons are reported per condition as counts.

\section{Results}

\begin{table}[t]
\centering
\footnotesize
\setlength{\tabcolsep}{3.5pt}
\caption{Pool B confirmatory factorial at the medium cap (136 tasks $\times$ 4 seeds; $n=544$ runs per cell). Var.\ is mean per-task outcome variance across seeds; Exh.\ counts terminations at the enforced cap only. Baseline and info-only runs instead hit the invisible 40k safety ceiling 136 and 121 times respectively.}
\label{tab:factorial}
\begin{tabular}{lrrrrr}
\hline
Cell & Res.\ \% & Tok/run & Tok/res. & Exh. & Var. \\
\hline
Baseline & 73.5 & 10{,}088 & 13{,}720 & 0 & .035 \\
Info-only & 76.5 & 9{,}281 & 12{,}136 & 0 & .039 \\
Silent-bare & 53.5 & 630 & 1{,}179 & 253 & .059 \\
Silent-placebo & 60.7 & 602 & 992 & 214 & .050 \\
Announced & 57.9 & 609 & 1{,}053 & 229 & .057 \\
\hline
\end{tabular}
\end{table}

\begin{table}[t]
\centering
\footnotesize
\setlength{\tabcolsep}{3.5pt}
\caption{Preregistered contrasts on Pool B confirmatory tasks: risk differences in percentage points, 95\% task-cluster bootstrap CIs, two-sided task-stratified permutation $p$. Bare abbreviates silent-bare; placebo abbreviates silent-placebo. Risk differences here and in Table~\ref{tab:arms} are computed before rounding of cell percentages.}
\label{tab:contrasts}
\setlength{\tabcolsep}{3pt}
\begin{tabular}{lrrr}
\hline
Contrast & RD & 95\% CI & $p$ \\
\hline
Announced $-$ placebo (primary) & $-2.8$ & $[-6.2, +0.7]$ & .11 \\
Announced $-$ bare & $+4.4$ & $[+0.7, +8.1]$ & .014 \\
Placebo $-$ bare & $+7.2$ & $[+4.0, +10.5]$ & .0002 \\
Info-only $-$ baseline & $+2.9$ & $[+0.7, +5.1]$ & .035 \\
Bare $-$ baseline & $-20.0$ & $[-25.0, -15.4]$ & .0002 \\
\hline
\end{tabular}
\end{table}

\begin{table}[t]
\centering
\footnotesize
\setlength{\tabcolsep}{3.5pt}
\caption{Supplementary arms. The tight-cap contrast reports the task-cluster bootstrap CI; contrasts are computed before rounding. Pool A dose-response rows pool silent and announced cells per cap level. Tight-cap cells are ordered bare / placebo / announced; perturbed and unperturbed rows are ordered baseline / placebo / announced. Rows without a pool label use Pool B confirmatory tasks. Perturbed tasks have renamed entry points; unperturbed values come from the same 46-task subsample within the confirmatory factorial.}
\label{tab:arms}
\setlength{\tabcolsep}{2pt}
\resizebox{\columnwidth}{!}{%
\begin{tabular}{lrr}
\hline
Condition & Value & $n$/cell \\
\hline
Pool A, tight cap (1 attempt, pooled) & 63.0 & 54 \\
Pool A, medium cap (2 attempts, pooled) & 70.8 & 72 \\
Pool A, generous cap (4 attempts, pooled) & 82.4 & 54 \\
Pool A medium: announced $-$ placebo & 0.0 & 72 \\
Pool A medium: bare $-$ baseline & $-23.6$ & 72 \\
Tight cap, per cell (\%) & 57.4 / 61.5 / 57.4 & 408 \\
Tight: announced $-$ placebo & $-4.2$ $[-8.1, -0.2]$ & 408 \\
Countdown / neutral line (\%) & 58.0 / 58.7 & 138 \\
Wordings: plan / conseq.\ / eff.\ (\%) & 56.5 / 54.3 / 58.7 & 138 \\
2nd model: uncapped / any cap (\%) & 11.1 / 0.0 & 36 / 108 \\
Perturbed (\%) & 59.8 / 34.8 / 38.0 & 92 \\
Unperturbed (\%) & 77.2 / 61.4 / 54.3 & 184 \\
Memorization (median sim.\ / flagged) & 0.317 / 8 of 164 & -- \\
\hline
\end{tabular}%
}
\end{table}

\begin{figure}[t]
\centering
\includegraphics[width=\columnwidth]{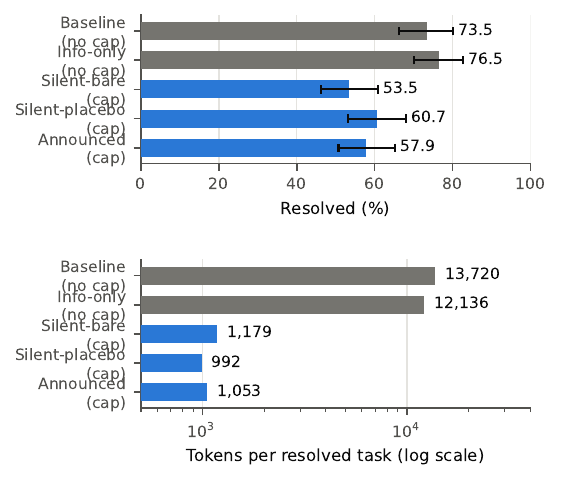}
\caption{Pool B confirmatory factorial at the medium cap ($n=544$ runs per cell). Top: resolved rate with 95\% task-cluster bootstrap intervals. Bottom: tokens per resolved task (ratio of sums, log scale). Gray: uncapped cells; blue: capped cells.}
\label{fig:factorial}
\end{figure}

\begin{figure}[t]
\centering
\includegraphics[width=\columnwidth]{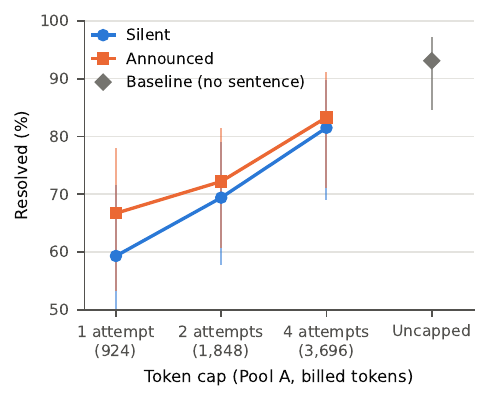}
\caption{Pool A dose-response. Resolved rate by cap level for silent and announced cells (tight and generous caps: $n=54$ per cell, three seeds; medium cap: $n=72$, four seeds; uncapped baseline: $n=72$, no disclosure arm). Whiskers are 95\% Wilson intervals over runs, unadjusted for task clustering and descriptive only; inference appears in Tables~\ref{tab:contrasts} and~\ref{tab:arms}.}
\label{fig:frontier}
\end{figure}

The primary contrast, answering RQ2, tests whether announcing an enforced budget changes task resolution relative to a placebo sentence at the same cap. On the 136 confirmatory Pool B tasks, the announced condition resolved 57.9\% of runs against 60.7\% for silent-placebo (Table~\ref{tab:factorial}, Figure~\ref{fig:factorial}), a risk difference of $-2.8$ percentage points (95\% CI $[-6.2, +0.7]$, $p=.11$; Table~\ref{tab:contrasts}). The confidence interval lies entirely inside the preregistered $\pm 10$-point equivalence bounds, establishing equivalence in the two-one-sided-tests sense at $\alpha=.05$, and tokens per resolved task were 1{,}053 for announced against 992 for placebo, a 6.1\% gap (Tables~\ref{tab:contrasts} and~\ref{tab:factorial}). These data establish that stating the budget produced no resolution or efficiency change beyond a content-free sentence at the medium cap.

The remaining factorial cells, answering RQ1 at the medium cap, isolate what enforcement does without any sentence and what a sentence does without enforcement. Silent-bare resolved 53.5\% of runs against 73.5\% for baseline, a difference of $-20.0$ points (95\% CI $[-25.0, -15.4]$, $p=.0002$), while tokens per resolved task fell from 13{,}720 to 1{,}179 (Tables~\ref{tab:factorial} and~\ref{tab:contrasts}). Announced exceeded silent-bare by $+4.4$ points (95\% CI $[+0.7, +8.1]$, $p=.014$), and silent-placebo exceeded silent-bare by $+7.2$ points (95\% CI $[+4.0, +10.5]$, $p=.0002$; Table~\ref{tab:contrasts}). In the two-by-two statement-by-enforcement decomposition, the interaction estimate was $+1.5$ points, computed on the same grid. The comparison establishes that enforcement produced the resolution loss and the efficiency gain, and that the announced condition's advantage over bare silence did not exceed the placebo's.

Supplementary conditions test whether the disclosure null is confined to the medium cap, the single upfront announcement, the primary wording, or memorized tasks. Across one-, two-, and four-attempt caps (tight, medium, generous), Pool A resolution averaged 63.0\%, 70.8\%, and 82.4\% (Table~\ref{tab:arms}, Figure~\ref{fig:frontier}). On Pool A's 18 confirmatory tasks at the medium cap, announced and silent-placebo resolved at the same rate (risk difference 0.0 points; Table~\ref{tab:arms}), while enforcement cost 23.6 points for a 4.6$\times$ efficiency gain, matching the Pool B pattern. The tight-cap arm was launched as a powered follow-up after the Pool A dose-response showed an exploratory $+7.4$-point disclosure advantage at its tight cap; the replication refuted the hint, with an announced-minus-placebo difference of $-4.2$ points (95\% CI $[-8.1, -0.2]$; Table~\ref{tab:arms}). That interval excludes zero and matches the medium-cap estimate in sign, so the data support a small resolution cost of disclosure, bounded below the smallest effect size of interest, and rule out any benefit approaching it. Per-turn countdown resolved 58.0\% against 58.7\% for a neutral injected line, and alternative wordings spanned 54.3--58.7\% against the primary wording's 57.9\% (Table~\ref{tab:arms}); no variant's point estimate approached the 10-point threshold, and only the medium- and tight-cap contrasts were powered to bound effects below it, which both did. Identifier renaming lowered absolute resolution (77.2\% to 59.8\% at baseline) while the cap effect persisted and the announced and placebo cells stayed within a few points of each other; excluding the eight memorization-flagged tasks changed no contrast (Table~\ref{tab:arms}).

Three secondary outcomes bound the conditions under which the primary null was measured. Answering RQ4, mean per-task outcome variance rose from 0.035--0.039 in uncapped cells to 0.050--0.059 in capped cells, roughly 1.5 times the uncapped level (Table~\ref{tab:factorial}). Info-only resolved $+2.9$ points above baseline in Pool B (95\% CI $[+0.7, +5.1]$, $p=.035$; Table~\ref{tab:contrasts}), and the secondary model resolved 11.1\% uncapped and 0\% under every cap, with all 108 capped runs budget-exhausted (Table~\ref{tab:arms}). Answering RQ3, budget sensitivity varied with cap level and model: resolution tracked cap level on the primary model in both pools, while the capability-constrained secondary model resolved nothing under any cap. These results qualify the primary claim: the disclosure null was measured under caps that raised outcome variance, on a model whose failures were budget-constrained; the secondary model shows the capability-constrained case.

Taken together, the factorial, its decomposition, the robustness variants, and the secondary outcomes describe one pattern across both pools. The pattern was measured on two function-level program-repair pools, open models served on CPU, and one-, two-, and four-attempt caps; no broader setting was tested. Under these conditions, the data establish that enforcing a token budget changed resolution, efficiency, and variance, while communicating it (upfront, per turn, or in any tested wording) added nothing beyond the effect of a sentence.

\section{Discussion}

Budget enforcement determines what a coding agent achieves under a token cap; budget communication does not. The review located the field's gap in budgets treated as configuration constants; manipulating enforcement and disclosure as crossed treatments closes it, because the crossed design separates what the cap causes from what the trajectory would have produced. The preregistered hypothesis that an announced cap yields equal or higher resolution with less waste was not confirmed: both point estimates ran opposite the hypothesized direction ($-2.8$ and $-4.2$ points), the tight-cap interval excluded zero, and the 10-point effect of interest was excluded throughout (Table~\ref{tab:contrasts}). H1, that tighter budgets lower expected resolution, was confirmed by the dose-response (Table~\ref{tab:arms}). The fact holds at one- and two-attempt caps on two repair pools, under a primary model resolving 73.5\% uncapped (Table~\ref{tab:factorial}).

Against the scaffold family, the dose-response qualifies the suspicion that raising an agent's budget limit is unlikely to increase performance~\parencite{Yang2024SWEagent}: under caps calibrated to bind, resolution tracked cap level from 63.0\% to 82.4\% (Table~\ref{tab:arms}). The efficiency-by-design family's premise, that constraining agent freedom purchases cost reductions at tolerable accuracy loss, is confirmed and priced: enforcement alone cut tokens per resolved task from 13{,}720 to 1{,}179 for 20.0 points of resolution~\parencite{Xia2024Agentless, Yao2023ReAct, Schick2023Toolformer}. The allocation family is unaffected in its own regime, since routing and compute-optimal scaling act before dispatch~\parencite{Panda2025Routing, Snell2024TestTime}, but its premise that budget level is the active lever now has within-trajectory causal support. The behavioral family is contradicted at its transfer point: specific, difficult targets did not outperform a content-free sentence in these agents~\parencite{Locke2002Goal}. The contradiction survives the theory's own moderator, because the countdown variant supplied progress feedback~\parencite{Locke2006Goal} and produced no gain (Table~\ref{tab:arms}). The efficient-reasoning finding that stated budgets compress single-turn reasoning~\parencite{Han2024TALE} is not contradicted: uncapped, the budget sentence coincided with modestly lower expenditure (Table~\ref{tab:factorial}), but the placebo comparison shows that under an enforced cap the budget's semantics moved nothing a content-free sentence did not.

The placebo cell produced the unexpected result: a content-free sentence raised resolution over bare silence by 7.2 points (Table~\ref{tab:contrasts}), while the budget sentence added nothing further. The most parsimonious explanation is prompt-composition sensitivity: any added system-prompt sentence shifts the conditioning distribution, and content is irrelevant at this scale. An alternative, that the placebo's evaluation-framing content itself changed behavior, cannot be ruled out, because a single placebo wording was tested. A second result ran opposite to the disclosure null: stating an unenforced cap raised resolution by 2.9 points at lower mean expenditure (Tables~\ref{tab:factorial} and~\ref{tab:contrasts}). The most parsimonious account is the same sentence effect with no cap to pay for it; a selection alternative (one of several secondary contrasts at $\alpha=.05$) cannot be excluded from these data.

\subsection*{Limitations and Threats to Validity}
The models were open weights served on CPU; under frontier API models with different instruction-following profiles the disclosure null may not hold; within deployments of comparable open models it holds as measured. All runs used one custom minimal repair harness; effects under standard tool-using scaffolds may differ, resolution is defined by hidden-test passes with pass-count-only feedback, and the models were 4-bit quantized. Because a quarter of baseline runs reached the invisible 40{,}000-token safety ceiling, the uncapped reference is itself weakly truncated, and the 20.0-point enforcement cost may understate the loss relative to truly unbounded runs. Tasks were function-level repairs with hidden tests; under repository-scale issues whose budgets fund exploration and navigation, both cap and disclosure effects may differ; for bounded repair tasks the estimates apply directly. Absolute resolution depended on identifier familiarity, falling from 77.2\% to 59.8\% under renaming (Table~\ref{tab:arms}); on fully novel code, levels would shift further; the contrasts, which survived renaming, are the transferable quantities. The design was powered for a 10-point effect; disclosure effects narrower than the reported intervals would escape detection; the null therefore bounds the effect below practical size and asserts nothing about exact zero.

For token-capped program repair, governance reduces to two priced decisions: whether to enforce a cap, and where to set it; the tested range moved resolution from 63.0\% to 82.4\% (Table~\ref{tab:arms}). The class of problems is wider than repair: any deployment in which an autonomous language-model system runs under a metered resource (tokens, tool calls, wall-clock time) and an operator chooses between silent and disclosed limits. Across that class, as tested here, the principle is that budget governance is enforcement design: prompt-level choices did not move outcomes. The placebo decomposition supports the principle's sharpest form: apparent gains from budget language are sentence effects, and the semantics of the budget contribute nothing detectable (Tables~\ref{tab:factorial} and~\ref{tab:contrasts}).

\section{Future Work}

The disclosure null was established against a single placebo wording, so the 7.2-point sentence effect is measured but not decomposed. The immediate next step is to rerun the capped cells with a panel of content-free sentences varying length, register, and topic, holding the cap at 1{,}532 tokens and the seed lists fixed. Constant resolution across the panel would confirm content-independent prompt-composition sensitivity; variation would locate which sentence properties carry the effect.

The estimates hold for function-level repairs with hidden tests; repository-scale issues, whose budgets fund exploration and navigation, were unreachable here because their evaluation requires container-based harnesses and budgets beyond this study's compute. The follow-up is the same five-cell factorial on repository-scale issue resolution~\parencite{Jimenez2024SWEbench}, with caps recalibrated from uncapped development percentiles and 40--60 preregistered tasks at four or more seeds. Transfer of the enforcement price and the disclosure null to that setting would extend both estimates to the tasks organizations actually govern; the harness, frozen design, and task-sampling configuration for this arm are included in the study archive as a ready-to-run replication package.

Enforcement-over-communication was established for token caps, a resource whose consumption agents cannot estimate reliably~\parencite{Bai2026Spend}. Interaction-round quotas~\parencite{Liu2024AgentBench} invert that property: remaining actions are exactly countable by the agent, so disclosure may carry usable information that token budgets cannot. Repeating the factorial under round quotas tests whether the principle concerns metered resources generally or only resources whose consumption the agent cannot track.

Before these estimates existed, organizations had no causal basis for dividing governance effort between enforcement design and communication design; the enforcement-only answer now makes the deployment question visible. Whether enforcement-only governance holds where humans supervise agents (overriding caps, resubmitting stopped runs, adjusting budgets over months) cannot be answered by benchmark harnesses and requires longitudinal field experiments in production engineering organizations. Such studies trade executable ground truth for organizational outcomes (cost predictability, override rates, delivery), a capability outside controlled repair benchmarks.

\section{Conclusion}

Token-budget governance for coding agents now has causal estimates: an enforced cap trades 20 percentage points of resolution for a roughly twelvefold reduction in tokens per resolved task, and announcing that cap changes nothing a placebo sentence does not. The blind choice named at the outset, where to set a limit and whether to tell the agent, no longer has to be blind, because both options now carry measured prices under executable verification. The measurements hold for function-level repair on open models at one-, two-, and four-attempt caps; they are estimates for that regime alone.

Platform engineers imposing token caps on coding agents can now set enforcement expecting a quantified resolution loss and efficiency gain; the trade is no longer unknown. The assumption that budget prompts require careful wording can be dropped: wording, timing, and repetition of budget language moved nothing that a content-free sentence did not move. Evaluation designers can likewise treat placebo sentence controls and preregistered enforcement contrasts as a justified template for assessing any harness-level prompt intervention. These prescriptions stop at the capability floor: on models that fail uncapped, caps filter tasks and no communication policy alters the outcome.

The placebo decomposition exposes a question the field could not previously ask: prompt interventions must beat a placebo sentence instead of an empty prompt, because a content-free sentence alone raised resolution by 7.2 percentage points over silent enforcement, a sentence effect that an empty-prompt control misattributes to the intervention. Whether the disclosure null survives frontier-scale agents on repository-level issues, where instruction following and planning exceed anything tested here, is the question that now determines if budget governance stays an enforcement-only problem. If the null holds there, governance effort belongs entirely in enforcement design; if it breaks, disclosure becomes a capability-dependent lever.

\section*{Acknowledgements}

Computational resources were provided by the computing facilities and cluster infrastructure of Louisiana Tech University under institutional access. The QuixBugs benchmark was accessed as an open-source release. The HumanEvalFix subset of HumanEvalPack was accessed as an open-access release.

\section*{Data and Artifact Availability}

The complete study archive, comprising the harness and evaluator source code, the frozen protocol and configuration, task splits and seed lists, per-run transcripts and token ledgers for all 5{,}835 runs, analysis scripts, and the ready-to-run cloud-arm replication package, accompanies this paper as supplementary material and will be archived in a DOI-issuing repository. QuixBugs and HumanEvalFix are used under their open-source licenses.

\section*{Declarations}

The authors declare no competing interests. The study involved no human subjects and no personal data; both task sets are publicly released benchmarks.
% Add a generative-AI-use disclosure sentence here per the target venue's policy.

\printbibliography

\end{document}

%% file: preamble.tex
\usepackage[letterpaper]{geometry}
\usepackage{paperstyle}
\usepackage{times}
\usepackage[none]{hyphenat}
\usepackage{url}
\usepackage{latexsym}
\usepackage{indentfirst}
\usepackage{graphicx}
\usepackage[
    style=apa,
  ]{biblatex}
\usepackage[hidelinks]{hyperref}